\documentclass[twocolumn,showpacs,pra,superscriptaddress]{revtex4}
\usepackage{dcolumn}
\usepackage{graphicx}
\usepackage{bbm}

\newcommand{\be}{\begin{equation}}
\newcommand{\ee}{\end{equation}}
\newcommand{\beq}{\begin{eqnarray}}
\newcommand{\eeq}{\end{eqnarray}}

\newcommand{\hide}[1]{}

\begin{document}

\title{Rydberg atom mediated polar molecule interactions: a tool for molecular-state conditional quantum gates
and individual addressability}

\author{Elena Kuznetsova}
\affiliation{Department of Physics, University of Connecticut,
Storrs, CT 06269}
\affiliation{ITAMP, Harvard-Smithsonian Center
for Astrophysics, Cambridge, MA 02138} 
\author{Seth T. Rittenhouse}
\affiliation{ITAMP, Harvard-Smithsonian Center
for Astrophysics, Cambridge, MA 02138}
\author{Hossein R. Sadeghpour}
\affiliation{ITAMP, Harvard-Smithsonian Center
for Astrophysics, Cambridge, MA 02138}
\author{Susanne F. Yelin}
\affiliation{Department of Physics, University of Connecticut,
Storrs, CT 06269} 
\affiliation{ITAMP, Harvard-Smithsonian Center
for Astrophysics, Cambridge, MA 02138}
\date{\today}

\begin{abstract}
We study the possibility to use interaction between a polar molecule in the ground electronic and vibrational state and a 
Rydberg atom to construct two-qubit gates between molecular qubits and to coherently control molecular states. A polar molecule within the electron orbit in a Rydberg atom can either shift the Rydberg state, or form Rydberg molecule. Both the atomic shift and the Rydberg molecule states
depend on the initial internal state of the polar molecule, resulting in molecular state dependent van der Waals or dipole-dipole interaction between Rydberg atoms. 
Rydberg atoms mediated interaction between polar molecules can be enhanced up to $10^{3}$ times. We describe 
how the coupling between a polar molecule and a Rydberg atom can be applied to coherent control of molecular states, specifically, 
to individual addressing of molecules in an optical lattice and non-destructive readout of molecular qubits.
\end{abstract}

\maketitle

\section{Introduction}

Ultracold polar molecules placed in periodic arrays of traps represent an
attractive system for quantum information processing \cite{DeMille} and
quantum simulation of condensed matter systems \cite{Baranov-review}. They
offer a variety of long-lived states for qubit encoding, described by various quantum 
numbers; rotational, spin and hyperfine (if electronic and nuclear 
spins are non-zero), or projections of electronic angular 
momenta $\Lambda$ (orbital) and $\Omega$ (orbital and spin) onto the 
molecular axis \cite{Carr}. Cold polar molecules also offer scalability to a large number of qubits. Polar molecules can be easily controlled by DC
electric and magnetic fields, as well as by microwave and optical fields
allowing to design various traps \cite{Traps,Andre}. A strong appeal of cold polar molecules,
however, is that they possess permanent electric dipole moments, allowing for long-range dipole-dipole interaction. The dipole-dipole
interaction offers a tool to construct two-qubit gates, required for
universal quantum computation \cite{DeMille,Susanne}. A naturally  scalable 
setup is an optical lattice, i.e. a periodic trapping potential produced by an optical standing wave formed from 
two counterpropagating laser beams. With a typical focus diameter of the beams $d \sim 1$ mm and 
the lattice period $\lambda/2 \sim 300-500$ nm, this system 
can accomodate $\sim (d/\lambda/2)^{2}\sim 10^{6}-10^{7}$ molecular qubits. A two-qubit gate, e.g. 
a phase gate ($\left| \epsilon_{1}\epsilon_{2}
\right\rangle\rightarrow e^{i\phi\epsilon_{1}\epsilon_{2}}\left|
\epsilon_{1}\epsilon_{2} \right\rangle$, where $\epsilon_{1,2}={0,1}$ are states of the first and second qubit), 
can be realized using dipole-dipole interaction when both molecules are in the qubit 
state $|1\rangle$ such that the two-qubit state $|11\rangle$
accumulates a phase $\phi=V_{\rm{int}}T=\pi$ during the interaction. 
Polar molecules that
can be currently produced at ultracold temperatures (required to place molecules in a lattice) 
by Feshbach and
photo-association are limited to alkali diatoms, having dipole moments of the
order of a Debye (1 Debye=$10^{-18}$ esu cm in CGS units). The resulting
dipole-dipole interaction strength scales as 
\begin{equation}
V_{\rm{int}}\sim \mu^{2}/R^{3} \sim 10^{-9}/R(\mathrm{cm})^{3}\;\mathrm{s}^{-1}
\end{equation}
Given the
gate duration $T\le 1$ $\mu$s, which is typically much shorter than the
qubit decoherence time, one finds that the required interaction strength can
be achieved at distances of the order of 
\begin{equation}  \label{eq:inter-distance}
R\sim (\mu^{2}T/\pi)^{1/3} \sim 100\;\mathrm{nm}
\end{equation}
In an optical lattice, these molecules interact efficiently with only
several nearest neighbours. To be able to take full advantage of the scalability offered by optical lattices, a mechanism has to be found to make 
molecules interact at greater distances, ideally at the opposite sides of the lattice.

The interaction strength can be significantly increased by coupling the state of a polar molecule to a Rydberg atom excitation (see Fig.\ref{fig:setup}a). 
One way that this can be achieved is by placing a polar molecule within a Rydberg electron's orbit where the ground state molecule can 
act as a perturber, shifting the electron energy. Alternatively the coupling can be realized through the creation of a new type of 
ultralong-range Rydberg molecule
proposed recently \cite{Seth}, where a
molecule is made of a polar molecule in the ground electronic state and a Rydberg atom. These Rydberg molecules can have extremely large dipole moments, on the order of a thousand Debye, comparable to dipole moments of Rydberg atoms. 
Moreover, the Rydberg molecules have different sizes and binding energies depending on the state of the molecular perturber. This means
that by coupling to Rydberg atoms, interaction between polar molecules can
be, first, strongly enhanced and, second, state-dependent. These two
conditions facilitate implementation of two-qubit gates between polar molecules
at large distances.

Not only Rydberg atoms can enhance the interaction, they can also provide
individual control over polar molecules. One realization of this
control would be individual addressing of molecules in an optical lattice, which still 
poses a problem.
It means that one qubit rotations and readout of molecular states can be
implemented without the need for experimentally challenging laser tight-focusing and single-site addressability. Another example would be the 
molecular control in small
registers, which form the basis of quantum communication with
quantum repeaters and distributed quantum computation. In summary, coupling
of polar molecules to Rydberg atoms offers a way to enhance the range of
efficient interaction as well as to realize coherent control over molecular
states, including individual
addressing of molecular qubits in a lattice.

The paper is organized as follows. In Section II we analyze two schemes
allowing to increase the efficient interaction range, based on van der
Waals (vdW) and dipole-dipole interaction between Rydberg molecules. We describe
how a two-qubit phase gate can be realized with KRb and CH molecules,
coupled to Rb atoms. In Section III, we discuss the possibility to
selectively perform one-qubit rotations and read out qubit states of polar
molecules by coupling  to Rydberg atoms. We also describe how a controlled-NOT gate
can be implemented between a polar molecule and a neutral atom mediated by an atomic
Rydberg state, which is the main ingredient of computation in quantum 
repeaters and small quantum registers.

\section{Enhancement of polar molecules interaction strength}

\begin{figure}[tbp]
\center{
\includegraphics[width=8.5cm]{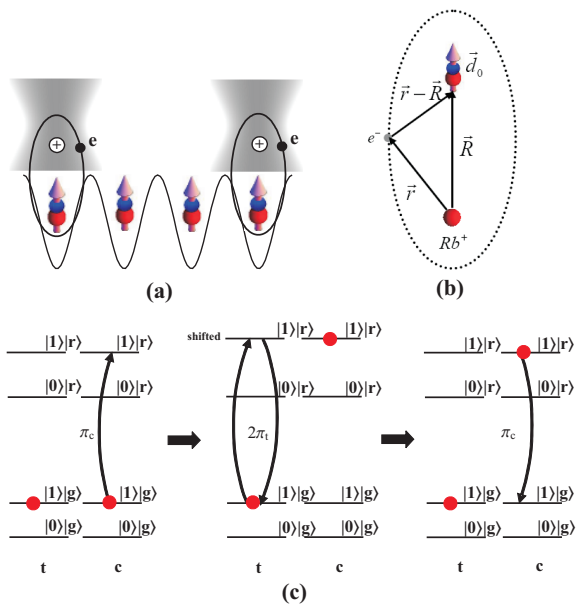}
\caption{\label{fig:setup} (a) Polar molecules in an optical lattice can be coupled with Rydberg atoms held in a separate lattice or individual optical 
dipole traps; (b) Coupling of a polar molecule to a Rydberg electron and a positive core. The molecule is separated from the core 
by $\vec{R}$, the electron-core distance is $\vec{r}$. The polar molecule is within the electron's orbit, schematically shown by a dotted ellips; 
(c) Two-qubit phase gate between polar molecules based on the Rydberg blockade mediated by atomic Rydberg-Rydberg interaction (see text for details)}
}
\end{figure}

We envision a setup shown in Fig.\ref{fig:setup}a, where polar molecules are
placed in an optical lattice while ground state neutral atoms are kept
either in an additional lattice or in individually controlled dipole traps.
Polar molecules made of alkali atoms can be formed in an optical lattice by
Feshbach associating pairs of atoms in each site and transferring the
resulting Feshbach molecules to the stable ground rovibrational state by
STIRAP (stimulated Raman adiabatic passage) \cite{Nagerl}. As we already mentioned in the introduction polar
molecules can interact efficiently with several neighbors via dipole-dipole
interaction. To enable interaction between two distant molecules, atoms are
brought sufficiently close to molecules so that the molecule is within a Rydberg electron orbit as is illustrated in Fig.\ref{fig:setup}b. A phase gate then can be realized using 
state-dependent dipole-dipole interaction (Rydberg blockade) \cite%
{blockade} in the following way: (i) a control atom is
conditionally excited (dependent on the qubit state of the control polar molecule) by a $\pi$-pulse to a Rydberg state $\left| r \right\rangle$; (ii) a $2\pi$-pulse of
the same wavelength is applied to an atom near the target polar molecule to excite it to
the same Rydberg state. The doubly excited state $\left| rr
\right\rangle$ is however shifted by the energy of the Rydberg-Rydberg vdW or
dipole-dipole interaction and is far off-resonance (the excitation is
blockaded). As a result, the $2\pi$-pulse has no effect on the target
molecule except for a small accumulated phase; (iii) the control atom is
deexcited back to the ground state by a second $\pi$-pulse. The
excitation sequence is shown schematically in Fig.\ref{fig:setup}c.

The blockaded phase gate can be realized provided that the interaction
strength $V_{\rm{int}}\gg \Omega_{2\pi}$, where $\Omega_{2\pi}$ is the Rabi
frequency of the $2\pi$-pulse. The Rabi frequency defines the speed of the
gate, since the total gate time $T_{\rm{gate}}=2\pi/\Omega_{\pi}+2\pi/\Omega_{2%
\pi}$ is a sum of durations of two $\pi$- and one $2\pi$-pulse. The gate
time has to be much shorter than a qubit decoherence time and the Rydberg lifetime ($T_{\rm Ryd}$). If a qubit is encoded in rotational states of a polar
molecule, qubit decoherence times $\sim 1$ ms can be achieved \cite{Andre},
limited by electric field fluctuations. Even longer qubit coherence times
(greater than a second) can be realized if electronic spin or hyperfine
states are used \cite{Our-PRA}, since these states are insensitive to electric field
fluctuations in the first order. For sufficiently highly-excited Rydberg states,
$T_{\rm Ryd} \sim 100-500$ $\mu$s are expected, setting the main limit on
gate duration and, therefore, on the Rabi frequencies. If we assume a gate duration $T_{\rm{gate}}\sim 1$ $\mu$s, short enough
compared to $T_{\rm Ryd}$, it will correspond to the Rabi
frequency $\Omega_{\pi}$, $\Omega_{2\pi}\sim 100$ kHz. The blockade radius,
which determines the range of efficient interaction, is estimated as $%
R_{b}=\left(C_{6}/\Omega\right)^{1/6}\sim (n^{11}/\Omega)^{1/6}$ (in a.u.) for vdW interaction.
For $n\sim 50$, $R_{b}\approx 5$ $\mu$m, about fifty times larger than
compared to direct interaction between polar molecules. Even larger blockade
radius of tens $\mu$m is possible if Rydberg molecules interact via
dipole-dipole interaction. In this case $R_{b} \sim (\mu^{2}/\Omega)^{1/3}$,
where $\mu\sim n^{2}$ is the dipole moment of a Rydberg molecular state.
Assuming again $n\sim 50$ and $\Omega \sim 100$ kHz, the resulting blockade
radius $R_{b}\approx 40$ $\mu$m. At distances larger than the blockade
radius the phase gate can be realized by simultaneously exciting both
control and target molecules to the Rydberg molecular state, letting them
interact via dipole-dipole interaction and accumulate a $\pi$ phase shift, and deexciting them back \cite{blockade}.
Therefore, extending the blockade radius from $R_b \sim 100$ nm to $R_b \sim 100$ $%
\mu$m allows molecules separated by $10^{3}$ lattice sites, to interact
instead of only few nearest neighbors.

A qubit can be encoded into long-lived rotational states of the ground electronic and
vibrational molecular state. Rotational transitions of polar molecules have
dipole moments equal to the permanent one and frequencies in the GHz
range, allowing for fast one-qubit gates with microwave pulses. Rotational
states are shifted by an electric field and as a result a qubit will
decohere in the presence of electric field fluctuations. The coherence time
can be significantly increased if a qubit is encoded into spin or hyperfine
sublevels and is swapped onto rotational states for one and two-qubit
operations. If we encode a qubit into the first rotational states $\left\vert
0\right\rangle ,\left\vert 1\right\rangle =\left\vert J=0,1,2\right\rangle $%
, where $J$ is the rotational angular momentum, we can selectively excite a
molecule plus an atom from a specific $J$ state to the Rydberg molecular state, in which molecules interact via 
dipole-dipole interaction. Another possibility is to couple the $J$ states to isotropic $ns$ Rydberg atomic
states, in which Rydberg atoms interact via vdW interaction.
Here we analyze KRb+Rb as an example system with vdW interaction between Rydberg atoms dependent on the state of polar 
molecules.

We may find the shift in the atomic Rydberg excitation frequency 
by considering the
shift of the polar molecule energy in a rotational state $J=0,1,2,...$due to
the presence of a Rb(ns) Rydberg atom. In effect, the polar molecule's
energy is shifted by the internal field of the positive core shielded by a
highly excited s-wave Rydberg electron. The internal electric fields of
a Rydberg atom are small (on the order of 10$^{-7}$ atomic units or several 100 V/cm) 
so that the field induced shift is small compared to the rotational splitting of the molecule (on the order of $\sim 10$ GHz). \
This means that there is very little mixing between different molecular rotational
states, and their energy shifts quadratically in the field strength. The shift of the 
rotational states can be found by diagonalizing a Hamiltonian whose matrix elements are \cite{Seth}
\begin{eqnarray}
H_{JJ^{\prime }} &=&B\frac{J\left( J+1\right) }{2}\delta _{J,J^{\prime
}}-\left\langle J\left\vert \vec{F}_{Ryd}\cdot \vec{d}_{0}\right\vert J^{\prime
}\right\rangle,   \label{Eq:Pol_mol_H} \\
\vec{F}_{Ryd} &=&e\hat{z}\left\langle \psi _{ns}\left\vert \frac{\cos
\theta _{\vec{R},\vec{r}}}{\left\vert R\hat{z}-\vec{r}\right\vert ^{2}}%
\right\vert \psi _{ns}\right\rangle,   \nonumber
\end{eqnarray}%
where $B$ and $\vec{d}_{0}$ is the polar molecule rotational constant and body
fixed permanent dipole moment. Here $\vec{F}_{Ryd}$ is the electric field at
the position of the polar molecule, $\vec{R}$, due to the
positive core of the Rydberg atom shielded by an $s$-wave electron, $\vec{r}$ is the position of the Rydberg electron
(see Fig.\ref{fig:setup}b), $\psi_{ns}$ is the Rydberg electron wavefunction.
\begin{figure}[tbp]
\center{
\includegraphics[width=9.cm]{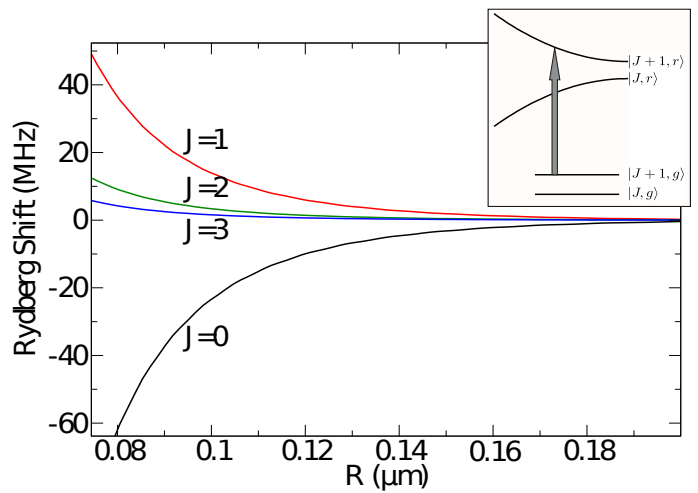}
\caption{\label{fig:Shifts}The shift in the excitation frequency of a Rb(50s) Rydberg excitation 
due to the presence of a KRb molecule in J =0, 1, 2, 3 states (labeled) is shown in terms of 
the core-molecule separation distance. The inset illustrates molecular state dependent excitation to an atomic $ns$ Rydberg state.}
}
\end{figure}

The predicted shift in the Rydberg excitation frequency due to a KRb
molecule perturber in several of the lowest rotational states is shown in Fig. \ref{fig:Shifts} as a
function of the core molecule separation, $R$. We have subtracted out the
zero field energies of the KRb rotational states as we are only interested
in the net shift in the Rydberg excitation energy. Due to the small internal
fields the mixing between different rotational states is very small (less
than 2\%), so even though the polar molecule is not in a pure $J$ state, we
have labeled each state by it's dominant angular momentum component. 
Fig. \ref{fig:Shifts} shows that for separation distances of the order of  100 nm, 
the shift in excitation frequency is in the range of several MHz to ten MHz. This shift can be resolved by a laser with a linewidth of
several kHz and a Rabi frequency of the order of $100$ kHz. The Rabi frequency of 
$100$ kHz results in a gate duration $T_{\rm{gate}}\sim 1 \mu s << T_{\rm Ryd}$. In a recent work, \cite{Ryd-mol-lifetime} a lifetime of a Rydberg molecular $5s_{1/2}-35s$ state of two Rb atoms has been 
measured $T_{\rm{Ryd}}=6.4$ $\mu$s. The lifetime is probably limited by collisions with ground state Rb atoms and is expected to be in the range of 
10--100 $\mu$s for an isolated Rydberg atom, which means that a gate time of $1$ $\mu$s is short enough compared to $T_{\rm Ryd}$.

The analysis presented above is appropriate for molecules with fairly small
dipole moments, less than 0.4 a.u. (1 Debye). It was shown in Ref. \cite{Rittenhouse2}
that for larger dipole moments the electron-dipole interaction is strong
enough to mix several higher electron angular momentum states which can
significantly change the energy landscape. Ref. \cite%
{Rittenhouse2} describes the potential energy surface that arises due to the
presence of a $\Lambda $-doublet molecular perturber (which can be modelled as a two-level system) with a larger dipole
moment. Describing the intricate 2D energy surface that arises in the case
of rigid rotor type molecules (having a multilevel structure of rotational states), 
similar to KRb or RbCs, is in progress.

As was shown in Fig.\ref{fig:Shifts} $ns$ states Rydberg atoms can interact via isotropic vdW
interaction, allowing for the effective interaction
range between polar molecules to be extended to several $\mu$m. A
larger interaction range can be realized using dipole-dipole
interaction between Rydberg molecules. As it has been shown, \cite{Seth}
states of a Rydberg atom with angular momenta $l\ge 3$ (those with
negligible quantum defects) are hybridized in the presence of a
$\Lambda$-doublet polar molecule perturber forming ultralong-range
Rydberg molecule states which can acquire dipole moments of several
kiloDebyes (kD). Furthermore, the vibrationally bound states formed in this
system correspond to a polar molecule perturber that is in an internal
state which is fully polarized with a dipole moment pointing either
toward or away from the positive Rydberg core. The
Rydberg molecule states for different internal configurations of the
polar molecule perturber are also separated spatially with potential
wells separated by 100s of nm.

This theory was further refined in \cite{Rittenhouse2}, where an
extended set of Rydberg atom states was considered by including all electron angular
momentum states for several principal quantum numbers $n$. For
molecules with permanent dipole moments larger than $d_0 \sim 0.4$
a.u., the resulting Rydberg molecules acquire significant s-wave electron contribution (30 - 40\%)
making the molecular state accessible by a standard two-photon Rydberg
excitation schemes. The resulting Rydberg molecule size and binding energy
is dependent on the internal orientation of the perturbing polar
molecule, and hence
conditioned on the initial  state of the polar molecule.
Excitation from a desired internal state can be tailored by tuning a
Rydberg excitation to the appropriate binding energy and by modifying
the spatial separation of the ground state atom from the polar
molecule.

\begin{figure}[tbp]
\center{
\includegraphics[width=7.5cm]{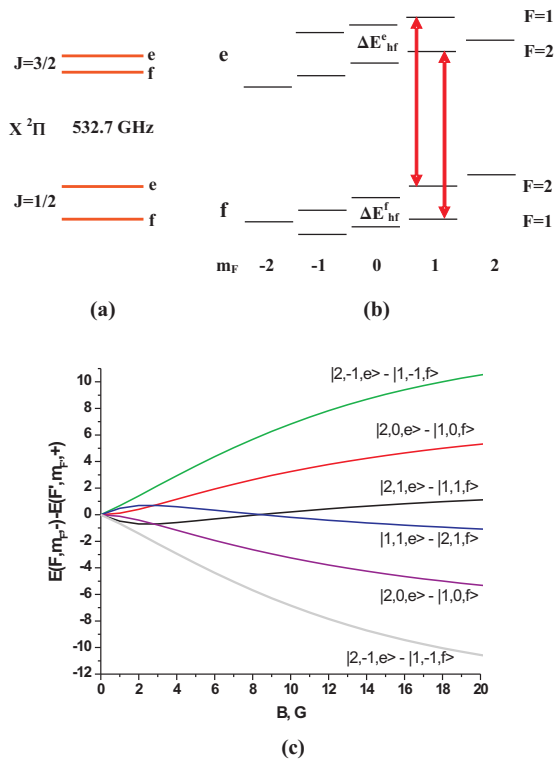}
\caption{\label{fig:CH} (a) Ground $^{2}\Pi$ state of CH split into $J=1/2,3/2$ states by spin-orbit interaction with $\sim 0.5$ THz splitting; 
(b) the hyperfine structure of the 
$^{2}\Pi_{3/2}$ state due to the hydrogen spin; (3) Zeeman shifts of the $\pi$ transition frequencies between satellite lines $|F=1,m_{F},e\rangle - |F=2,m_{F},f\rangle$, 
$|F=2,m_{F},e\rangle - |F=1,m_{F},f\rangle$, $m_{F}=1$ transitions are magnetic field insensitive at $B=2.4$ G and can be used for qubit encoding. }
}
\end{figure}

$\Lambda$-doublet molecules with two opposite parity states $\left(f,e\right)$ far separated from higher rotational states fit this analysis particularly 
well. One good example is CH with a $^{2}\Pi_{3/2}$ metastable state having lifetime $>10^{3}$ s \cite{CH-lifetime}, split into $\Lambda$-doublet $\left(f,e\right)$ states with a $700$ MHz 
splitting, as shown in Fig.\ref{fig:CH}a. 
Each $\Lambda$-doublet state is further split into two hyperfine sublevels due to the hydrogen nuclear spin (see Fig.\ref{fig:CH}b). Therefore, it offers an 
opportunity to encode a qubit into hyperfine sublevels $|F,m_{F},p\rangle$, where $p={f,e}$ is the parity of the state. As was observed in the ground state OH (which has a similar $\Lambda$-doublet and hyperfine structure)
in \cite{OH-qubits}, two combinations of hyperfine states, namely, $|F=2,m_{F}=1,e/f\rangle$ and $|F=1,m_{F}=-1,e/f\rangle$ are insensitive 
to fluctuations of a magnetic field in the first order at $B=3.01$ G and $B=7.53$ G, respectively, similar to hyperfine clock 
states $|F,m_{F}=0\rangle$ and $|F=2,m_{F}=1\rangle$, $|F=1,m_{F}=-1\rangle$ in Rb atoms. The insensitivity to magnetic field fluctuations makes these 
states good candidates for qubit encoding.

We carried out similar analysis for the $^{2}\Pi_{3/2}$ state of CH. The $\left(f,e\right)$ hyperfine states 
shift in the magnetic field as 
\begin{eqnarray}
E^{f,e}=-\frac{\Delta E_{hf}^{f,e}}{2(2J+1)}+\mu_{B}g_{J}^{f,e}m_{F}B\pm \nonumber \\
\frac{|\Delta E_{hf}^{f,e}|}{2}\sqrt{1-\frac{4\mu_{B}g_{J}^{f,e}m_{F}B}{\Delta E_{hf}^{f,e}(2J+1)}+\left(\frac{\mu_{B}g_{J}^{f,e}B}{\Delta E_{hf}^{f,e}}\right)^{2}}  
\end{eqnarray}
where the $\pm$ sign before the square root refers to $F=J\mp 1/2$ for the $e$, and to $F=J \pm 1/2$ for the $f$ state; 
$\Delta E_{hf}^{e}=-20.908$ MHz and $\Delta E_{hf}^{f}=2.593$ MHz are the hyperfine splittings of the $f,e$ states \cite{CH-hyperfine} (the hyperfine structure in the 
$e$ state is inverted), $g_{J}^{f}=0.819537$ and $g_{J}^{e}=0.817829$ \cite{Timur} are the slightly different g-factors of the $f,e$ states, $\mu_{B}$ is the 
Bohr magneton and $B$ is the magnetic field. We found two transitions, $|F=1,m_{F}=1,e\rangle - |F=2,m_{F}=1,f \rangle$ and $|F=2,m_{F}=1,e\rangle - |F=1,m_{F}=1,f\rangle$ 
that become insensitive to the magnetic field at $B=2.4$ G (see Fig.\ref{fig:CH}c). The $g_{J}$-factor values are correct up to $\Delta g_{J}\approx 0.05 g_{J}$, but this variation does 
not change the results appreciably.  
The splitting of the CH $\Lambda$-doublet is $700$ MHz and the 
dipole moment for the transition between the $\left(f,e\right)$ states is $1.47$ D \cite{Brown}, allowing to apply fast one-qubit rotations with microwave pulses.

Two-photon excitation to the high dipole 
moment states of a Rydberg molecule, which are composed of $l> 2$ angular momentum states of a Rydberg atom can be realized due to fortunate 
overlap of $n(l>2)$ and $(n+\nu)s$ states, where $\nu$ is some number of the order of a quantum defect for the $s$ state ($\nu=3$ for Rb). This overlap 
mixes the $nl$ and $(n+\nu)s$ states which result in the mixed state wavefunction 
\begin{equation}
\psi=a(d_{0})\psi_{d}(\vec{r})|\xi_{d}\rangle+b(d_{0})\psi_{s}(\vec{r})|\xi_{s}\rangle,
\end{equation}
where $\psi_{d}$ and $|\xi_{d}\rangle$ are the electronic wave function and internal polar molecule state, that include higher electron angular 
momentum with $l>2$; $\psi_{s}$ and $|\xi_{s}\rangle$ are the s-wave Rydberg electron wave function and the corresponding molecular state, respectively. 
The expansion coefficients $a(d_{0})$ and $b(d_{0})$ depend on the dipole moment of the polar perturber as well as its position $R$. The $|a(d_{0})|^{2}$ 
determines the dipole moment of the state scaling as $d_{\rm{Ryd}}\approx 1.3|a(d_{0})|^{2}n^{2}$, which can be in kD range ( for $n=25$ 
and $a(d_{0})\approx 0.6$ the dipole moment $d_{\rm{Ryd}}\approx 1.4$ kD) \cite{Rittenhouse2}. On the other hand, the sizable $b(d_{0})$ coefficient would allow a convenient 
two-photon excitation from the ground atomic state. While for a two-state molecule, such as a $\Lambda$-doublet molecule, this theory works quite 
well, for a rigid rotor molecule, such as KRb, the formalism has to be extended to take into account the dependence of the polar molecule dipole interaction 
with the Rydberg atom electric field on the orientation of the rotating dipole. While we expect the results to be similar, it requires further investigation.

Finally, we note that the bound states of the Rydberg molecule form at different  $R$ and correspond to particular orientations 
of the polar molecule dipole moment. Namely, two dipole moment orientations, toward and from the positive core, result in two molecular wells, left and 
right, having smaller and larger molecule-core separation $R$. It means that molecules cannot be directly excited from pure $\Lambda$-doublet states $(f,e)$ to the 
Rydberg molecule bound states, since in the $(f,e)$ states a molecule does not have any specific dipole orientation. To realize the selective 
excitation, a Hadamard gate can be first applied to the qubit states, forming superpositions $|L,R\rangle =(|f\rangle \pm |e\rangle)/\sqrt{2}$. The left and 
right states $|L,R\rangle$ can now be selectively coupled to the left and right wells of a Rydberg molecule.  

\section{Coherent control of polar molecules by coupling to Rydberg atoms}

Interaction between a polar molecule and a Rydberg atom opens the
possibility to coherently control molecular states. In particular, we
will show below that individual addressing and readout can be realized
without resorting to tightly focused laser beams. This is similar to the
idea of a "marker" atom for individual manipulation of neutral atoms in a
lattice (atom microscope) \cite{Tommaso,Atom-microscope}. In our case the
Rydberg atom mediated coherent control realizes a "molecule microscope".

Individual addressing of atoms and molecules in an optical lattice is a
long-standing problem due to the difficulty of focusing laser beams to a
single lattice site. Recently, individual addressing of single atoms in a 
lattice has been achieved by tightly focusing a laser beam to a single site 
in a spin-dependent lattice \cite{individ-address}. An
alternative approach proposed in \cite{Tommaso} is to use "marker" atoms
that can selectively interact with atoms in a lattice. "Marker" atoms, held
in a separate optical lattice, can be brought to the same lattice site with
addressed atoms and interact via state-dependent collisions. Qubit states
acquire interaction-induced shifts, which offer a way to address the atom
by applying microwave or Raman pulses resonant with the shifted qubit
states. In the same spirit, addressing can be realized for polar molecules by
coupling them to Rydberg atoms. The main difference is that the interaction
is not relying on atom-molecule collisions (which can be inelastic or result
in a chemical reaction), but on a long-range interaction. Rotational states
of a polar molecule will be shifted due to the interaction with the Rydberg atom,
which will allow to perform one-qubit gates individually. 

\begin{figure}[tbp]
\center{
\includegraphics[width=8.5cm]{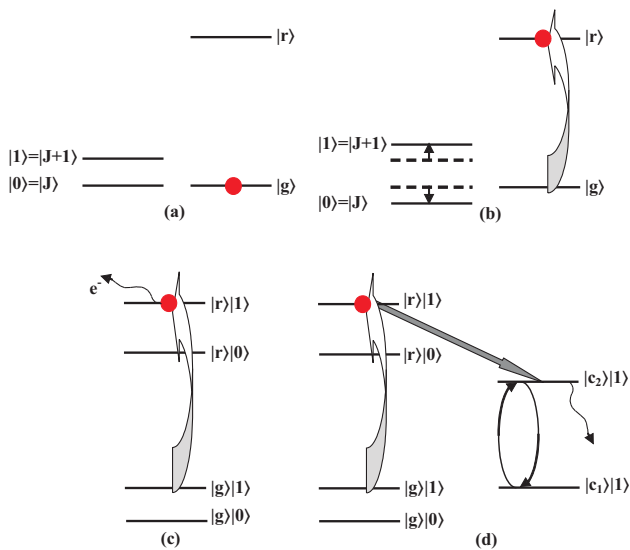}
\caption{\label{fig:schemes} (a) and (b): a scheme of individual adressing of molecules by coupling to Rydberg atoms. Once an atom is excited to the 
Rydberg state, rotational states shift, allowing to address the molecular qubit by a resonant microwave pulse; (c) individual readout of molecular 
states by selective excitation to a Rydberg state followed by ionization; (d) non-destructive readout by selectively transferring the system to a state where 
the atom undergoes cycling excitation between states $|c_{1,2}\rangle$ and detecting resonant fluorescence.}
}
\end{figure}

It is illustrated
in Fig.\ref{fig:schemes}a and b: when an atom is in the ground electronic
state there is no interaction between the atom and the molecule, and
molecular rotational states are not shifted. Once the atom is excited to the
Rydberg state, molecular states shift and the qubit transition frequency
changes. A microwave pulse resonant to the shifted frequency will act on the
addressed qubit without affecting the rest. As was shown in Fig.\ref{fig:Shifts} at distances 
of $\sim 100$ nm between a polar molecule and a Rydberg atom the rotational states shift by several MHz. 
A  microwave pulse resonant to the shifted qubit transition with a Rabi frequency 
of $\sim 100$ kHz, much smaller than the shift, will selectively interact with the addressed molecule. The resuling 
one-qubit gate time is again in the $\mu$s range, much shorter than the Rydberg state lifetime $T_{\rm{Ryd}}$. 
Similarly, molecular qubits can
be read out by molecular state dependent excitation of a Rydberg atom,
followed by e.g. ionization of the Rydberg state (see Fig.\ref{fig:schemes}%
c). We note that qubit-carrying molecules are not destroyed in the process,
although the Rydberg atom gets ionized and has to be replaced. Completely
non-destructive readout maybe possible if after the selective excitation to
the Rydberg molecular state the atom can be transferred to a state that is a
part of a cycling transition (Fig.\ref{fig:schemes}d). The readout is then
realized by detecting resonant fluorescent photons. Once a photon is detected, both
the molecule and the atom can be recycled.

Coherent coupling between Rydberg atoms and polar molecules can be utilized
to build small quantum registers, which form the basis of quantum repeaters 
\cite{QRepeater,QRepeater-Misha} and distributed quantum computing \cite%
{Distrib-QC}. Let us briefly summarize the idea of a quantum repeater as was
originally proposed in \cite{QRepeater}. In quantum communication to teleport a quantum
state a Bell state has to be shared between a sender
$A$ and a reciever $B$. The fastest and most efficient way proposed so far has been to send a pair of
entangled photons. Photons, however, cannot propagate large distances
without loss and depolarization. A quantum
repeater can help to entangle distant parties $A$ and $B$ by using intermediate nodes $C_{i}$%
, $i=1..N$ to reduce losses and depolarization errors. In the first step
pairs of nearest nodes get entangled: A with $C_{1}^{1}$,  $C_{1}^{2}$ with $%
C_{2}^{1}$, $C_{i}^{2}$ with $C_{i+1}^{1}$, etc., where $C_{i}^{1,2}$ are
the two (which is the minimal number of qubits per node required to implement
the protocol) qubits at the $i^{th}$ node. This initial nearest
neighbor entanglement is realized by measuring photons emmited by
communication qubits of each pair in the Bell basis. In the second step nodes $A$ and $C_{L}$, $C_{L}$ and $C_{2L}$, etc.
get connected by measuring intermediate qubit pairs $C_{i}^{1,2}$ at each
node in the Bell state basis. This is followed by entanglement purification in the obtained entangled pairs
to compensate for local one and two-qubit and measurement errors.
In the third step nodes $A$ and $C_{L^{2}}$, $C_{L^{2}}$ and $C_{2L^{2}}$,
etc. are connected in the same way. The process is repeated until $A$ is
connected to $B$ directly without any nodes in between. A key ingredient in connecting two entangled pairs 
of qubits into a single one and in entanglement purification is the ability to perform a CNOT gate between two 
qubits at each node. To illustrate how the CNOT gate is used in entanglement purification, we
describe as an example an entanglement swapping between
two pairs of qubits. Suppose we have two qubits (qubits 1 and 2) in a $\left| \Phi^{+}
\right\rangle=(\left| 00 \right\rangle+\left| 11 \right\rangle)/\sqrt{2}$
Bell state and two qubits (qubits 3 and 4) in a $\left| 0 \right\rangle$ state, and we want to
swap the entaglement between the two pairs. To this end, first a CNOT is applied to the qubits in $\left| 0 \right\rangle$
state controlled by the qubits forming the entangled pair 
\begin{eqnarray}
(\left| 00 \right\rangle^{12}+\left| 11 \right\rangle^{12})\left| 00
\right\rangle^{34}/\sqrt{2}{\quad\stackrel{1\rightarrow 3,2\rightarrow 4}{\longrightarrow}\quad}\nonumber \\
(\left| 00 \right\rangle^{12}\left| 00 \right\rangle^{34}+\left| 11
\right\rangle^{12}\left| 11 \right\rangle^{34})/\sqrt{2},
\end{eqnarray}
where qubits $1,2$ control qubits $3,4$ respectively. As a next step, qubits $1,2$ are measured in the $\left| \pm
\right\rangle=(\left| 0 \right\rangle\pm \left| 1 \right\rangle)/\sqrt{2}$
basis with the measurement outcomes $\left| ++ \right\rangle^{12}$, $\left| --
\right\rangle^{12}$ producing $|\Phi^{+}\rangle^{34}$, while measurement outcomes $\left| +- \right\rangle^{12}$ and $%
\left| -+ \right\rangle^{12}$ producing the $|\Phi^{-}\rangle^{34}$ Bell states. 

\begin{figure}[tbp]
\center{
\includegraphics[width=8.5cm]{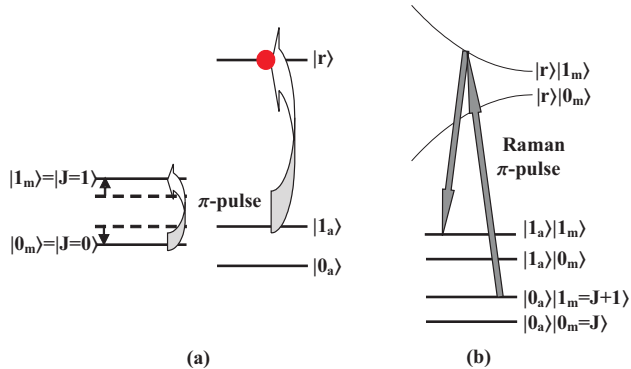}
\caption{\label{fig:CNOT}Controlled NOT gate between a polar molecule and a ground state atom mediated by Rydberg atomic states: 
(a) NOT gate on a molecular qubit controlled by the state of the atom. Exciting the atom from a specific qubit state, e.g. $|1_{a}\rangle$  
to a Rydberg state $|r\rangle$ and bringing the atom close to the molecule shifts the rotational states. A microwave $\pi$ pulse resonant with the shifted 
rotational transition flips the molecular qubit; (b) NOT gate on an atomic qubit controlled by a molecular qubit. The neutral atom qubit can be 
flipped depending on the state of the molecular qubit, e.g. $|1_{m}\rangle$,  by applying a Raman $\pi$ pulse resonant to the shifted state of a 
molecule-Rydberg atom state $|r\rangle |1_{m}\rangle$.}
}
\end{figure}

A ground state polar molecule and a neutral atom can form a quantum register with 
the molecule serving as a memory and the atom that can serve as a communication qubit to create 
initial entangement between nearest nodes and for
measurement. The quantum repeater protocol therefore can be implemented in this system provided a CNOT operation 
can be realized on a
polar molecule controlled by the atomic qubit and vise versa. In Fig.\ref{fig:CNOT} we illustrate 
how this can be done using intermediate Rydberg atomic states. A conditional flip of the
molecular qubit can be realized using the slightly modified scheme of Fig.%
\ref{fig:schemes}a,b: we selectively excite the atom to a Rydberg state only
from the qubits state $\left| 1_{a} \right\rangle$ and bring it at a distance $R$ of efficient 
interaction with the polar molecule, which will shift the
molecular qubit frequency. The molecular qubit state then can be flipped by a microwave $\pi$ pulse 
resonant to the shifted qubit frequency. If initially the atom was in the 
$\left| 0_{a} \right\rangle$ state it will not be excited to the Rydberg
state, the molecular qubit states will not shift and the $\pi$-pulse will be
off-resonant. The result is the molecular qubit flip if the atom is in the $%
\left| 1_{a} \right\rangle$ state (see Fig.\ref{fig:CNOT}a). A conditional flip of the atomic qubit can
be realized by a Raman $\pi$-pulse resonant to the atomic Rydberg state,
corresponding to a specific state of the molecular qubit, e.g. the molecular 
$\left| 1_{m} \right\rangle$ state as shown in Fig.\ref{fig:CNOT}b.

\section{Conclusions}

We propose and study interactions between ground state polar molecules and  Rydberg atoms as a tool in quantum information processing with polar 
molecules in optical lattices. We show that this interaction results in a molecular state dependent shift of $ns$ Rydberg atom states, which can be used 
to enhance the interaction range between molecules by many folds. The interaction range can be enhanced by excitation of ground state atoms to Rydberg 
states conditioned on the molecular state and allowing them to interact via long-range vdW interaction, increasing the interaction range up to several 
$\mu$m. Even larger interaction range of tens $\mu$m can be realized if the molecule and the Rydberg atom form a Rydberg molecule. Rydberg molecular 
states are expected to have dipole moments of several kD, allowing for strong long-range dipole-dipole interaction. The size of a Rydberg molecule as 
well as the energies of the bound vibrational states depend on the state of the polar molecule, resulting in molecular state-dependent interaction. 
We illustrate our proposal using rigid rotor KRb+Rb and $\Lambda$-doublet CH+Rb as examples of vdW and dipole-dipole interactions in Rydberg states. 
We also analyze the possibility to use the molecule-atom coupling to coherently control the state of the polar molecule with applications to individual 
addressing and molecular state readout as well as building small quantum registers composed of a polar molecule and a neutral atom.

\section{Acknowledgements}

The authors thank Timur Tscherbul for calculating g-factors for the $^{2}\Pi_{3/2}$ state of CH and many fruitful discussions. The work was supported 
by the AFOSR under the MURI award FA9550-09-1-0588 and NSF via a grant to ITAMP at the 
Harvard-Smithsonian Center for Astrophysics and Harvard Physics Department.


\begin{thebibliography}{99}
\bibitem{DeMille} D. DeMille, Phys. Rev. Lett. \textbf{88}, 067901 (2002).

\bibitem{Baranov-review} M. A. Baranov, Phys. Rep. \textbf{464}, 71 (2008).

\bibitem{Carr} L. D. Carr, D. DeMille, R. V. Krems, J. Ye, New J. Phys. 
\textbf{11}, 055049 (2009).

\bibitem{Traps} M. Lysebo, L. Veseth, Phys. Rev. A {\bf 83}, 033407 (2011).

\bibitem{Andre} A. Andre et al., Nat. Phys. \textbf{2}, 636 (2006).

\bibitem{Susanne} S. F. Yelin, K. Kirby, R. C\^ot\'e, Phys. Rev. A \textbf{74%
}, 050301 (2006).

\bibitem{Seth} S. T. Rittenhouse, H. R. Sadeghpour, Phys. Rev. Lett. \textbf{%
104}, 243002 (2010).

\bibitem{Nagerl} J. G. Danzl, M. J. Mark, E. Haller, M. Gustavsson, R. Hart,
J. Aldegunde, J. M. Hutson, H. C. Nagerl, Nat. Phys. \textbf{6}, 265 (2010).

\bibitem{blockade} D. Jaksch, J. I. Cirac, P. Zoller, S. L. Rolston, R.
C\^ot\'e, M. D. Lukin, Phys. Rev. Lett., \textbf{87}, 037901 (2001).

\bibitem{Our-PRA} E. Kuznetsova, R. Cote, K. Kirby, S. F. Yelin, Phys. Rev. A {\bf 78}, 012313 (2008). 

\bibitem{Ryd-mol-lifetime} B. Butscher, J. Nipper, J. B. Balewski, L. Kukota, 
V. Bendkowsky, R. L\"ow, T. Pfau, Nat. Phys. {\bf 6}, 970 (2010).

\bibitem{Rittenhouse2} S.T. Rittenhouse, M. Mayle, P. Schmelcher and H.R.
Sadeghpour, arXiv:1101.5353v1 (2011).

\bibitem{CH-lifetime} M. C. McCarthy, S. Mohamed, J. M. Brown, P. Thaddeus, Proc. Nat. Acad. Sc. {\bf 103}, 12263 (2006).

\bibitem{OH-qubits} B. L. Lev, E. R. Meyer, E. R. Hudson, B. C. Sawyer, J. L. Bohn, J. Ye, Phys. Rev. A {\bf 74}, 061402 (2006).


\bibitem{CH-hyperfine} T. Amano, Astrophys. J. {\bf 531}, L161 (2000).

\bibitem{Timur} T. V. Tscherbul, private communication.

\bibitem{Brown} J. M. Brown, A. Carrington, {\it Rotational Spectroscopy of Diatomic Molecules}, Cambridge University Press, Cambridge (2003). 



\bibitem{Tommaso} T. Calarco, U. Dorner, P. S. Julienne, C. J. Williams, and
P. Zoller, Phys. Rev. A \textbf{70}, 012306 (2004).

\bibitem{Atom-microscope} A. Klinger, S. Degenkolb, N. Gemelke, K.-A.
Brickman Soderberg, C. Chin, Rev. Sci. Instr. \textbf{81}, 013109 (2010).

\bibitem{individ-address} C. Weitenberg, M. Endres, J. F. Sherson, M.
Cheneau, P. Schaub, T. Fukuhara, I. Bloch, S. Kuhr, arxiv:1101.2076 (2011).

\bibitem{QRepeater} H.-J. Briegel, W. D\"ur, J. I. Cirac, P. Zoller, Phys.
Rev. Lett. \textbf{81}, 5932 (1998); L. Childress, J. M. Taylor, A. S.
Sorensen, M. D. Lukin, Phys. Rev. A \textbf{72}, 052330 (2005).

\bibitem{QRepeater-Misha} L. M. Duan, M. D. Lukin, J. I. Cirac, P. Zoller,
Nature \textbf{414}, 413 (2001).

\bibitem{Distrib-QC} E. T. Campbell, Phys. Rev. A \textbf{76}, 040302
(2007); L. Jiang, J. M. Taylor, A. S. Sorensen, M. D. Lukin, Phys. Rev. A 
\textbf{76}, 062323 (2007).
\end{thebibliography}
\end{document}